\documentclass[pre,twocolumn,amsmath,amssymb,floatfix,longbibliography]{revtex4}
\usepackage{mathrsfs,bm}
\usepackage{longtable}
\usepackage{txfonts}
\usepackage{indentfirst}
\usepackage{graphicx,color,dcolumn,booktabs}
\usepackage{multirow}
\usepackage{indentfirst}
\usepackage{multirow}
\usepackage{epsfig}
\usepackage{graphicx,color,dcolumn}
\usepackage{epstopdf}
\usepackage{amsmath}
\usepackage{multirow}
\usepackage{diagbox}
\usepackage{changes}
\usepackage{threeparttable}
\usepackage{enumitem,overpic,chemarr}
\usepackage[english]{babel}
\usepackage{color}
\usepackage{amssymb}
\definecolor{cover}{rgb}{0.77,0.87,0.88}
\definecolor{blueone}{rgb}{0.1,0.1,.7}
\definecolor{citec}{rgb}{0.14,0.47,0.09}
\definecolor{two}{rgb}{0.0,0.5,0.}
\definecolor{three}{rgb}{.5,.1,0.15}
\usepackage[bookmarks=true,bookmarksopen=false,plainpages=false,breaklinks=true,
   bookmarksnumbered=true,hypertexnames=false,
   filecolor=blue,urlcolor=three,menucolor=three,
   linkcolor=three,citecolor=blueone, colorlinks,
   anchorcolor=blue,runcolor=pink,frenchlinks=red
   pdfstartview=FitH,pdftitle=title,%
   pdfauthor=author]{hyperref}

\usepackage{relsize}
\usepackage{xspace}
\def\babar{\mbox{\slshape B\kern-0.1em{\smaller A}\kern-0.1em
    B\kern-0.1em{\smaller A\kern-0.2em R}}}

\allowdisplaybreaks[4]
\usepackage{color}
\usepackage{graphicx}
\usepackage{dcolumn}
\usepackage{bm}

\usepackage[utf8]{inputenc}
\usepackage[T1]{fontenc}
\usepackage{mathptmx}
\usepackage{etoolbox}
\allowdisplaybreaks[4]

\usepackage{color}
\begin{document}

\title{A Novel Nonlinear IP$_3$R State Transition Model and Calcium Oscillation}
\author{Zhao-Yu Peng,  Han-Yu Jiang\footnote{Corresponding author:jianghy@njnu.edu.cn}, Jun He\footnote{Corresponding author: junhe@njnu.edu.cn}}
\affiliation{$^1$School of Physics and Technology, Nanjing Normal University, Nanjing 210097, China}

\date{\today}

\begin{abstract} We present a novel nonlinear state transition model for inositol
1,4,5-trisphosphate receptors (IP$_3$Rs) that incorporates a pre-activated state,
as suggested by electron microscopy observations. Our model provides a
theoretical framework for the biphasic Ca$^{2+}$ dependence of IP$_3$Rs and accurately
reproduces their experimentally observed state distribution under saturating IP$_3$
conditions. By integrating receptor dynamics with cytoplasmic and endoplasmic
reticulum (ER) calcium exchange, we simulate IP$_3$R-mediated Ca$^{2+}$ oscillations
governed by six key conformational states. A pivotal finding is that IP$_3$
regulates these oscillations in a switch-like manner: once a critical IP$_3$
concentration is reached, the system abruptly transitions to sustained,
constant-amplitude oscillations that quickly terminate when the concentration
exceeds a secondary threshold. These results underscore the crucial role of the
pre-activated state in modulating calcium signaling.


\end{abstract}

\maketitle

\section{INTRODUCTION}\label{sec1}

IP$_3$R is a key calcium channel in the ER that mediates intracellular Ca$^{2+}$
release, thereby shaping spatiotemporal Ca$^{2+}$ signals that regulate diverse
cellular processes~\cite{berridge_inositol_2016}. Ca$^{2+}$ functions as a
versatile second messenger, and its signals, such as puffs, waves, and
oscillations, regulate critical processes including transcription, mitochondrial
activity, and vesicular trafficking. The frequency, amplitude, and spatial
distribution of Ca$^{2+}$ signals determine specific physiological outcomes and
are essential for maintaining cellular homeostasis.

Beyond fundamental cellular functions, IP$_3$R-mediated Ca$^{2+}$ signaling is
crucial for autophagy, cell proliferation, differentiation, and apoptosis.
Dysregulation of these processes has been implicated in various diseases,
including cancer, cardiovascular disorders, neurodegenerative conditions, and
metabolic diseases~\cite{vicencio_inositol_2009,kania_ip3_2017}. Abnormal
Ca$^{2+}$ oscillations can disrupt gene expression, alter mitochondrial
metabolism, and impair apoptotic mechanisms, contributing to disease progression
and cellular dysfunction. Understanding the regulatory mechanisms of IP$_3$R is
therefore essential for elucidating disease pathophysiology and developing
targeted therapeutic strategies~\cite{santulli_intracellular_2017,
mikoshiba_role_2015, cui_targeting_2017, berridge_calcium_2012}.

Early models of IP$_3$R-mediated Ca$^{2+}$ oscillations, based on reaction
kinetics, described intracellular Ca$^{2+}$ dynamics and wave propagation but
did not explicitly incorporate discrete conformational transitions of
IP$_3$R~\cite{Dupont_1994, atri1993single, duffy1998traveling,
kupferman1997analytical}. State transition models provide a mechanistic
framework for describing IP$_3$R function, explicitly accounting for transitions
between different channel states~\cite{dupont_calcium_2011}.  Various
deterministic models have been proposed, some simplifying IP$_3$R as a single
functional unit, while others explicitly consider its tetrameric nature and
cooperative subunit interactions~\cite{Dupont_1994, falcke_reading_2004,
Thul_2007, swillens1999calcium, siekmann2012kinetic}. Among these, the
DeYoung-Keizer (DYK) model remains widely used, offering a kinetic framework
that captures key regulatory features of IP$_3$R gating and Ca$^{2+}$
oscillations~\cite{de_young_single-pool_1992}. It describes both
Ca$^{2+}$-induced activation at low Ca$^{2+}$ concentrations and
Ca$^{2+}$-dependent inhibition at high concentrations.  At the molecular level,
IP$_3$R consists of four subunits, each containing an IP$_3$ binding site and
multiple Ca$^{2+}$ binding sites that cooperatively regulate channel
activity~\cite{foskett_inositol_2007}. Ca$^{2+}$ binding at high-affinity sites
promotes activation, while binding at low-affinity inhibitory sites leads to
inactivation. Meanwhile, IP$_3$ binding enhances receptor sensitivity,
facilitating channel opening. Once activated, IP$_3$R mediates Ca$^{2+}$ release
from the ER into the cytoplasm, amplifying the signal through Ca$^{2+}$-induced
Ca$^{2+}$ release. As cytosolic Ca$^{2+}$ levels rise, a balance between
Ca$^{2+}$-induced activation and Ca$^{2+}$-dependent inactivation generates
oscillatory Ca$^{2+}$ dynamics~\cite{Schmitz_2022, paknejad_structural_2018,
berridge_inositol_2016}. These transitions form the foundation of IP$_3$R
function, making state transition models essential for understanding the
regulation of Ca$^{2+}$ signaling.

Building on the DYK model, Li and Rinzel employed the fast-slow time-scale separation method to reduce the system’s complexity while preserving key dynamical features of the original model~\cite{li1994equations}. Further refinements were introduced by Shuai et al., who developed several simplified models by modifying state transitions in the DYK model to better align with experimental observations ~\cite{shuai_kinetic_2007, shuai_investigation_2009}. These models simulated key channel properties, including the open probability of IP$_3$R, the mean open time ($\tau_0$), and the mean closed time across varying Ca$^{2+}$ concentrations, validating the model’s predictions against patch-clamp recordings from Xenopus laevis oocytes. To better capture the interplay between intracellular organelles, subsequent models have incorporated mitochondrial Ca$^{2+}$ uptake and release dynamics~\cite{qi_optimal_2015,kiviluoto_regulation_2013, bartok_ip3_2019}.

Despite significant progress,  a key challenge in modeling these channels lies in determining which states in the model corresponds to a physiologically meaningful step in the activation process of IP$_3$R. Traditional models often assume a direct transition between inactive, active, and inhibited states. However, recent cryo-EM studies reveal that IP$_3$R undergoes additional intermediate conformational changes, challenging these simplified assumptions.
Paknejad et al. conducted Ca$^{2+}$ titration experiments on human IP$_3$R and obtained cryo-EM structures across five orders of magnitude of Ca$^{2+}$ concentration~\cite{paknejad_structural_2023}. Their findings revealed that Ca$^{2+}$ binding to the activating site does not immediately trigger channel opening. Instead, the receptor first transitions into a previously unidentified pre-activated state before reaching full activation. This intermediate state suggests that IP$_3$R activation is a multistep process, where conformational rearrangements must occur before the channel can fully open.

While some pioneering models, such as the nine-state model by Shuai et al., divided the activated state in the DYK model into distinct pre-activated and activated states, recent cryo-EM data reveal a more complex sequence of transitions that are not fully captured by existing mathematical models~\cite{shuai_kinetic_2007}. Notably, these new structural insights suggest that the pre-activated state serves as an obligatory intermediate before the receptor reaches its fully activated form. However, experimental observations indicate that the activated state, rather than the pre-activated state, undergoes inhibition. This discrepancy underscores the need for a refined model that explicitly integrates the newly identified pre-activated state and accurately reflects its role in the sequential activation and inhibition of IP$_3$R, thereby providing a more precise framework for studying the mechanisms driving IP$_3$R-mediated Ca$^{2+}$ oscillations.

To address these limitations, we propose a nonlinear IP$_3$R state transition
model that explicitly incorporates the pre-activated state, in alignment with
recent structural findings. By refining the representation of IP$_3$R gating
dynamics, our model captures the sequential conformational transitions leading
to channel activation. Additionally, by integrating Ca$^{2+}$ flux between the
cytoplasm and ER, the model offers new insights into the regulation of
IP$_3$R-mediated Ca$^{2+}$ oscillations.

The article is organized as follows: Section~\ref{sec2} presents our nonlinear
IP$_3$R state transition model, incorporating a pre-activated state and
Ca$^{2+}$ flux between the cytoplasm and ER. In
Section~\ref{sec3}, we validate the model against human IP$_3$R state
distribution and simulate IP$_3$R-dependent Ca$^{2+}$ oscillations nad , identifying
six key conformational states and an IP$_3$ concentration threshold for channel
activation. Section~\ref{sec4} concludes the article.

\section{Model}\label{sec2}

Unlike the RyR calcium channel, which is often described using a few discrete
states, such as in the Keiser-Levine model~\cite{Keizer_1996}, also applied in
our previous works on calcium oscillations and sparks~\cite{Gao_2023, Li_2025},
the IP$_3$R channel is typically modeled as a tetramer. Each subunit contains
three key ligand-binding sites: one for IP$_3$, one for Ca$^{2+}$ activation,
and one for Ca$^{2+}$ inhibition~\cite{bartok_ip3_2019, fan_cryo-em_2018}. Each
of these sites can be either unoccupied (0) or occupied (1), leading to eight
possible states per subunit. In the original DYK
model~\cite{de_young_single-pool_1992}, these eight states are defined by their
ligand-binding status and represented as a cubic model, where each vertex
corresponds to a state and each edge represents a transition. Initially, each
subunit starts in the 000 state, with no ligands bound. Classical models
typically designate the 110 state as active and the 111 state as inhibited.

\begin{figure}[h] \includegraphics[bb=0 0 1900 1850, clip,scale=0.11]{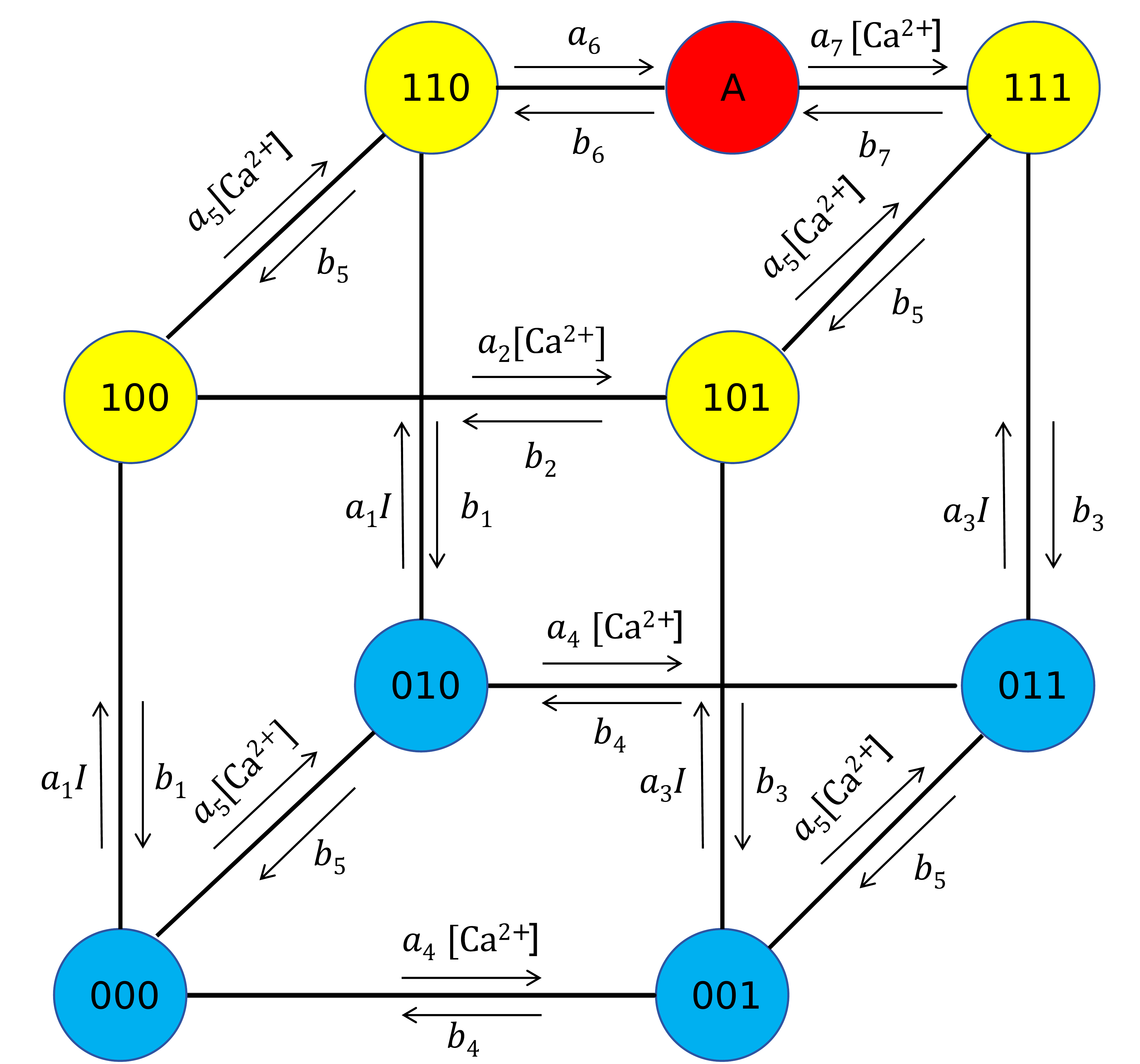} \caption{Nine-state model of the IP$_3$R channel classified by binding status. Blue-filled circles represent states where IP$_3$ is unbound, yellow-filled circles indicate states where IP$_3$ is bound, and red-filled circles denote the fully activated state A. The 110 state is the pre-activated state. Arrows represent state transitions, with transition rates influenced by IP$_3$ and $\text{Ca}^{2+}$ concentrations.} \label{fig:model} \end{figure}

Recent cryo-EM studies suggest the presence of an additional intermediate state before full activation, highlighting the need for a refined model~\cite{paknejad_structural_2023}. As shown in Fig.~\ref{fig:model}, we introduce a novel nine-state IP$_3$R channel model that incorporates conformational changes observed in electron microscopy studies of IP$_3$R activation~\cite{paknejad_structural_2023}. In this model, upon Ca$^{2+}$ binding at the activation site, the subunit transitions into the 110 state, representing a pre-activated state. After a brief dwell time, it reaches the fully activated state A, opening the channel and allowing Ca$^{2+}$ to be released into the cytoplasm. As local Ca$^{2+}$ concentrations rise, Ca$^{2+}$ binds to the inhibitory site, shifting the channel into the 111 state, leading to closure.

By incorporating structural insights from cryo-EM studies, our model offers a refined perspective on IP$_3$R channel dynamics. Earlier models, such as the nine-state model by Shuai et al.~\cite{shuai_kinetic_2007}, distinguished between pre-activated and activated states but were developed without structural data. In these models, the activated state A was directly connected only to the pre-activated 110 state, with inhibition occurring at the 110 state. However, cryo-EM studies reveal that the pre-activated state is an obligatory intermediate before full activation and, importantly, that inhibition occurs only after activation, rather than at the pre-activated stage~\cite{paknejad_structural_2023}.  Consequently, in our model, the activated state is positioned between the pre-activated 110 state and the inhibited 111 state.

The transition rates between states depend on the concentrations of signaling molecules and their associated rate constants, as shown in Fig.~\ref{fig:model}. In this model, we assume that Ca$^{2+}$ binding follows diffusion-limited kinetics, leading to a linear dependence on cytosolic [Ca$^{2+}$]. This is justified by the fact that physiological Ca$^{2+}$ concentrations (typically 0.001–10~$\mu$M) are well below the saturation levels of most binding sites, where binding is governed by diffusional encounter rates. Such linear dependence has also been widely adopted in classical models of IP$_3$R, including the DYK model and its extensions~\cite{de_young_single-pool_1992,shuai_kinetic_2007, shuai_investigation_2009}. The specific rate constants for these transitions are listed in Table~\ref{tab:parameters}.
\renewcommand\tabcolsep{0.205cm}
\renewcommand{\arraystretch}{2}
\begin{table}[h!]
\caption{Parameter values used in the nine-state model of IP$_3$R. Parameters highlighted in bold are newly introduced in our model and are fitted based on data from Ref.~\cite{paknejad_structural_2023}. All other parameters are adopted from the DYK model~\cite{de_young_single-pool_1992}. The symbol $\delta r$ denotes the change in the Pearson correlation coefficient resulting from a 50\% variation of the corresponding parameter.}
\begin{tabular}{c l l l l}
\toprule[1pt]
Parameter & Value  & Unit & $\delta r$ & Description \\ 
\hline
$a_1$ & $400$&$\mu$M$^{-1}$s$^{-1}$&0.006& IP$_3$(binding) \\ 
$a_2$ & $0.2$&$\mu$M$^{-1}  s^{-1}$ &0.002& Ca$^{2+}$(binding) \\
$a_3$ & $400$&$\mu$M$^{-1} s^{-1}$ &0.030& IP$_3$(binding) \\
$a_4$ & $0.2$&$\mu$M$^{-1} s^{-1}$ &0.001& Ca$^{2+}$(binding) \\
$a_5$ & $20$&$\mu$M$^{-1}s^{-1}$ &0.094& Ca$^{2+}$(binding) \\
$a_6$ & $\bf 80$&$s^{-1}$ &0.111& Ca$^{2+}$(activation) \\
$a_7$ & $\bf 0.2$&$\mu $M$^{-1} s^{-1}$ &0.089& Ca$^{2+}$(inhibition) \\
$b_1$ & $52$ &s$^{-1}$ &0.004& IP$_3$(unbinding) \\
$b_2$ & $0.2098$&s$^{-1}$ &0.031& Ca$^{2+}$(unbinding) \\
$b_3$ & $377.36$&s$^{-1}$ &0.019& IP$_3$(unbinding) \\
$b_4$ & $0.0289$&s$^{-1}$ &0.006& Ca$^{2+}$(unbinding) \\
$b_5$ & $1.6468$&s$^{-1}$ &0.070& Ca$^{2+}$(unbinding) \\
$b_6$ & $\bf 100$&s$^{-1}$&0.093&  Ca$^{2+}$(inactivation) \\
$b_7$ & $\bf 0.2098$& s$^{-1}$ &0.040& Ca$^{2+}$(activation) \\ 
\bottomrule[1pt]
\end{tabular}
\label{tab:parameters}
\end{table}

Most of the parameters in our model are adapted from the classical DYK model, which provides well-established rate constants for Ca$^{2+}$ and IP$_3$ binding and unbinding (parameter pairs $a_1$–$b_1$, $a_2$–$b_2$, ..., $a_5$–$b_5$)~\cite{de_young_single-pool_1992}. These parameters primarily determine the affinities of the activation and inhibition sites and have been widely used in previous IP$_3$R modeling studies. For the newly introduced transition rates governing the pre-activated to conducting states ($a_6$–$b_6$ and $a_7$–$b_7$),  no direct experimental measurements are available for these rates.  We determine $a_6$–$b_6$ and $a_7$–$b_7$ by minimizing the discrepancy—quantified via the Pearson correlation coefficient $r$—between the model-predicted and experimentally observed state distributions across a range of fixed Ca$^{2+}$ concentrations under saturated IP$_3$ conditions, as revealed by cryo-electron microscopy~\cite{paknejad_structural_2023}. This structure-based fitting ensures that the model captures gating dynamics consistent with observed conformational states. Moreover, the fitted parameters are further validated by demonstrating that the model reproduces physiologically relevant calcium oscillations, thereby linking structural data with functional calcium signaling behavior.

The resulting fit yields a Pearson correlation of $r = 0.628$ between simulated and experimental state distributions. To evaluate the sensitivity of the model, we examined how a 50\% variation in each parameter affects the Pearson correlation coefficient $r$. For completeness, we also report the sensitivity results for the fixed parameters, to provide a broader context.
The resulting change in correlation, denoted by $\delta r$, indicates that variations in $a_5$, $a_6$, $a_7$, $b_5$, and $b_6$ each lead to a change in $\delta r$ of approximately 0.1, suggesting these parameters have a substantial impact on model performance. In contrast, variations in the remaining parameters result in significantly smaller changes, with $\delta r$ consistently below 0.05.

Early models, such as the DYK model, assumed that although IP$3$R is a tetramer, only three equivalent and independent subunits contribute to conduction. In this simplified framework, the channel was considered open only when all three subunits were in the activated state ( which is also the  ${110}$ state, as the preactivated and activated states were not differentiated in these models), leading to an open probability of $P_{\rm O} = x_{A}^3$~\cite{de_young_single-pool_1992, li1994equations}. However, in reality, IP$_3$R functions as a tetramer, with each monomer's activation regulated by both IP$_3$ and Ca$^{2+}$. Recent experimental and theoretical studies tend to favor the idea that subunit interactions play a crucial role, and the channel opens only when at least three of the four subunits are in the activated state~\cite{shuai_kinetic_2007, shuai_investigation_2009, qi_optimal_2015}. Therefore, the open probability is given by
\begin{align} P_{\rm O}=P_{\rm 4O} + P_{\rm 3O} = x_{A}^4 + 4 x_{A}^3 (1 - x_{A}). \end{align}

To simulate calcium oscillations, it is crucial to account for calcium exchange between the cytosol and the ER. In this study, we adopt the simplified framework of the DYK model, which has been widely used in formulating the differential equations governing calcium dynamics in both compartments \cite{de_young_single-pool_1992, foskett_inositol_2007, marhl_complex_2000, means_reaction_2006}. The calcium concentration in the cytoplasm evolves according to the differential equation
\begin{align}
\frac{{\rm d}[\text{Ca}^{2+}]}{{\rm d}t} = J_{\text{in}} - J_{\text{out}},
\label{de}
\end{align}
where $[\text{Ca}^{2+}]$ represents the free $\rm{Ca}^{2+}$ concentration in the cytoplasm. 
Calcium influx primarily occurs through two mechanisms: IP$_3$R-mediated release and passive leakage, which together contribute to
\begin{align}
J_{\text{in}} = c_1 (\nu_1 P_{\rm O} +\nu_2) \left([\text{Ca}^{2+}]_{\text{ER}} - [\text{Ca}^{2+}]\right).
\end{align}
Here, $[\text{Ca}^{2+}]_{\text{ER}}$ is the $\rm{Ca}^{2+}$ concentration in the
ER, and $c_1$ denotes the volume ratio between the ER and the cytoplasm. The
first term, $\nu_1 P_{\rm O}$, represents $\rm{Ca}^{2+}$ release through the
IP$_3$R channel, where $\nu_1$ is the maximum flux and $P_{\rm O}$ denotes the
channel open probability. The second term, $\nu_2$, accounts for passive
$\rm{Ca}^{2+}$ leakage through non-gated channels.  To ensure conservation of
total $\rm{Ca}^{2+}$ within the system, we impose the constraint
$c_0=[\text{Ca}^{2+}]_{\text{ER}}+[\text{Ca}^{2+}]$.

In our model, we assume that the maximum flux is identical for states with either three or four activated subunits. Currently, there is no direct experimental evidence linking the number of activated subunits to distinct unitary flux levels. Although single-channel recordings have revealed multiple conductance states, the precise relationship between these states and subunit activation remains unclear~\cite{Mak_1994}. Alzayady et al. demonstrated that full channel activation under physiological conditions requires all four subunits to be activated~\cite{alzayady_defining_2016}, whereas the classical DYK model assumes that three activated subunits suffice for channel opening. Nonetheless, several modeling studies allow both 3/4- and 4/4-activated states to be conductive, often assuming comparable maximal flux~\cite{shuai_kinetic_2007,Greene_2024}. We adopt this modeling convention as it simplifies the analysis and avoids introducing parameters currently unconstrained by experimental data.

The primary mechanism for transporting $\rm{Ca}^{2+}$ from the cytoplasm back into the ER is the SERCA calcium pump. This calcium efflux is governed by a second-order Hill equation,
\begin{align}
J_{\text{out}} = \frac{\nu_3 [\text{Ca}^{2+}]^2}{[\text{Ca}^{2+}]^2 + k_3^2},
\end{align}
where $\nu_3$ represents the maximum calcium uptake rate of the SERCA pump, and $k_3$ is the activation constant. The key parameters used in these equations are summarized in Table~\ref{para}.
\renewcommand\tabcolsep{0.08cm}
\renewcommand{\arraystretch}{1.5}
\begin{table}[h!]
\caption{Parameters for the exchange of calcium between the cytosol and the ER \cite{de_young_single-pool_1992}.}
\label{para}
\begin{tabular}{l l l}
\toprule[1pt]
Parameter & Value & Description \\ 
\hline
$c_0$ & 2.0 $\mu$M & Total [Ca$^{2+}$] in terms of cytosolic vol \\ 
$c_1$ & 0.185 & (ER vol)/(cytosolic vol) \\ 
$\nu_1$ & 6 s$^{-1}$ & Max Ca$^{2+}$ channel flux \\ 
$\nu_2$ & 0.11 s$^{-1}$ & Ca$^{2+}$ leak flux constant \\ 
$\nu_3$ & 0.9 $\mu$M$^{-1}$s$^{-1}$ & Max Ca$^{2+}$ uptake \\ 
$k_3$ & 0.1 $\mu$M & Activation constant for ATP-Ca$^{2+}$ pump \\ 
\bottomrule[1pt]
\end{tabular}
\end{table}

\begin{figure*}[htpb]
  \centering
  \includegraphics[bb=-10 0 920 400, clip,scale=0.555]{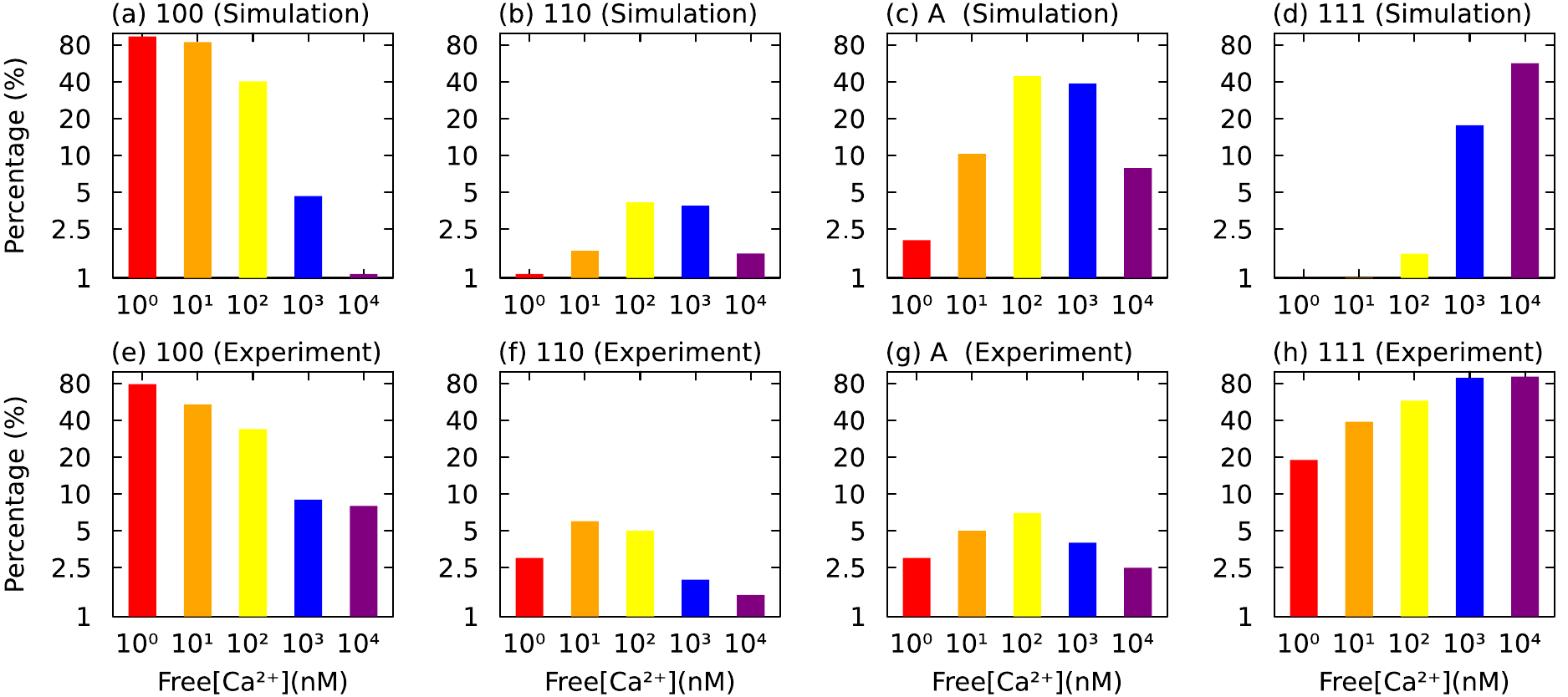} 
\caption{State distribution at different fixed Ca$^{2+}$ concentrations under saturated IP\textsubscript{3} condition. The patterns (a, b ,c, d) depict the simulated state distributions of the 100, 110, A, and 111 states at varying orders of magnitude of Ca$^{2+}$ concentration, respectively. The patterns (e, f, g, h) show the corresponding state distributions reploted  from the results of electron microscopy experiments~\cite{paknejad_structural_2023}. Both the simulations and experiments were conducted with IP$_3$ = 2 $\mu$M.}
  \label{fig: states distribution}
\end{figure*}

\section{RESULTS}\label{sec3}

\subsection{methodology}
We developed an IP$_3$R state transition model that incorporates calcium exchange between the cytosol and the endoplasmic reticulum (ER). This model aims to reproduce the state distribution of human IP$_3$R reported by Paknejad et al.~\cite{paknejad_structural_2023}. By coupling the IP$_3$R model with calcium cycling dynamics, we simulate IP$_3$R-dependent Ca$^{2+}$ oscillations and explore the functional roles of different IP$_3$R states.

To obtain the equilibrium state distributions under varying calcium ion concentrations ([Ca$^{2+}$]), we employed a nine-state transition rate matrix framework. We defined a state probability vector $P(t)$ and constructed a $9 \times 9$ transition rate matrix $K$, which captures all possible transitions between states. Off-diagonal elements $K(j,i)$ represent transition rates from state $i$ to state $j$, while diagonal elements $K(i,i)$ correspond to the total rate of leaving state $i$. These transitions mainly reflect IP$_3$ and Ca$^{2+}$ binding/unbinding events, making transition rates dependent on ligand concentrations, as well as ligand-independent conformational changes.

The steady-state distribution was obtained by numerically solving the equation $K \cdot P = 0$ with normalization of $P$. This calculation was performed using Julia’s DifferentialEquations.jl package, leveraging its robust solvers for linear systems. We computed the state distributions over a physiologically relevant range of calcium concentrations (100 to 10,000 nM), enabling a detailed representation of how Ca$^{2+}$ modulates IP$_3$R state probabilities. Incorporating the flux expression from Eq.~(\ref{de}), we then simulate calcium oscillation dynamics by solving the resulting nonlinear system.

 \subsection{Distribution of IP\textsubscript{3}R States under saturated IP\textsubscript{3} condition}

Electron microscopy studies were conducted under conditions of saturated
IP\textsubscript{3}, with several fixed Ca\textsuperscript{2+}
concentration levels~\cite{paknejad_structural_2023}. To mirror these experimental
conditions, we perform simulations with 2 $\mu$M IP\textsubscript{3}, while
maintaining Ca\textsuperscript{2+} concentrations at 1, 10,
10\textsuperscript{2}, 10\textsuperscript{3}, and 10\textsuperscript{4} nM,
following the experimental methodology in Ref.~\cite{paknejad_structural_2023}.
Since the Ca\textsuperscript{2+} concentration remained constant, calcium
exchange between the cytosol and the ER is not considered. The state
distribution is determined solely based on the nine-state IP$_3$R model. Given
that IP\textsubscript{3} is saturated, only states where IP$_3$ is bound are
relevant. We compare the distributions of four key states (100, 110, A, and 111)
with experimental electron microscopy data, as shown in Fig.~\ref{fig: states
distribution}.

To quantitatively assess the agreement between our simulated and experimentally observed state distributions, we compute the Pearson correlation coefficient between the predicted and measured state distribution. The resulting correlation coefficient $r = 0.628$, indicating a moderate-to-strong agreement. While this level of correlation is not perfect, it reflects that the model captures the key trends in the relative state occupancies.

It is important to note that the experimental state distributions derived from cryo-EM data are inherently subject to multiple sources of uncertainty, including classification noise during 3D reconstruction, limited particle numbers, and biological heterogeneity across IP$_3$R tetramers. Furthermore, variations in calcium buffering and sample preparation may also influence the measured conformational occupancies. Considering these factors, a correlation of this magnitude  supports the reliability of our model in capturing the essential behavior of the IP$_3$R gating landscape.

Our results reveal a distinct Ca\textsuperscript{2+}-dependent distribution of
IP\textsubscript{3}R states. At low Ca\textsuperscript{2+} concentrations ($<$10
nM), the majority of IP\textsubscript{3}Rs reside in the 100 state, where the
receptor is bound to IP\textsubscript{3} but remains unbound to
Ca\textsuperscript{2+}. As Ca\textsuperscript{2+} concentration increases,
IP\textsubscript{3}Rs gradually transition to the pre-activated 110 state, where
one Ca\textsuperscript{2+} ion is bound. This pre-activated state serves as an
intermediate, facilitating subsequent activation into the A state. The relative
abundance of the activated A state reaches its peak (43.4\%) when
[Ca\textsuperscript{2+}] is within 10\textsuperscript{2} nM. However, at very
high Ca\textsuperscript{2+} concentrations ($>$10\textsuperscript{4} nM), the
receptor undergoes inhibition, with most molecules transitioning into the 111
state, in which both Ca\textsuperscript{2+}-binding sites are occupied. These
trends in the four states presented here closely align with electron microscopy
data, as shown in the lower four patterns of Fig.~\ref{fig: states
distribution}~\cite{paknejad_structural_2023}. 

The distributions of individual receptor states exhibit characteristic trends
across the full range of Ca\textsuperscript{2+} concentrations. As expected, the
occupancy of the 100 state decreases with increasing [Ca\textsuperscript{2+}],
while the 111 state increases. In contrast, both the pre-activated 110 and
activated A states follow a biphasic pattern: their relative abundances
initially increase with [Ca\textsuperscript{2+}], reaching a peak before
declining at higher concentrations. This bell-shaped distribution is consistent
with previous electrophysiological recordings and theoretical
studies~\cite{de_young_single-pool_1992,Chen_2016,foskett_inositol_2007,Bezprozvanny_1991},
confirming that IP\textsubscript{3}Rs exhibit a well-defined biphasic activation
profile.

\subsection{Open Probability at Different IP\textsubscript{3}R Concentrations and the Role of the IP\textsubscript{3} State}

To assess the functional significance of different states in our model, we systematically removed the 001, 101, and 010 states from the original nine-state IP\textsubscript{3}R channel model, generating three simplified versions. Using these models, we conducted numerical simulations to analyze the opening probability $P_{\rm O}$ as a function of $[\text{Ca}^{2+}]$ concentration under varying IP\textsubscript{3} levels, considering only IP\textsubscript{3}R state transitions.

\begin{figure*}[htbp]
\centering
\includegraphics[bb=0 0 3200 1700, clip, scale=0.161]{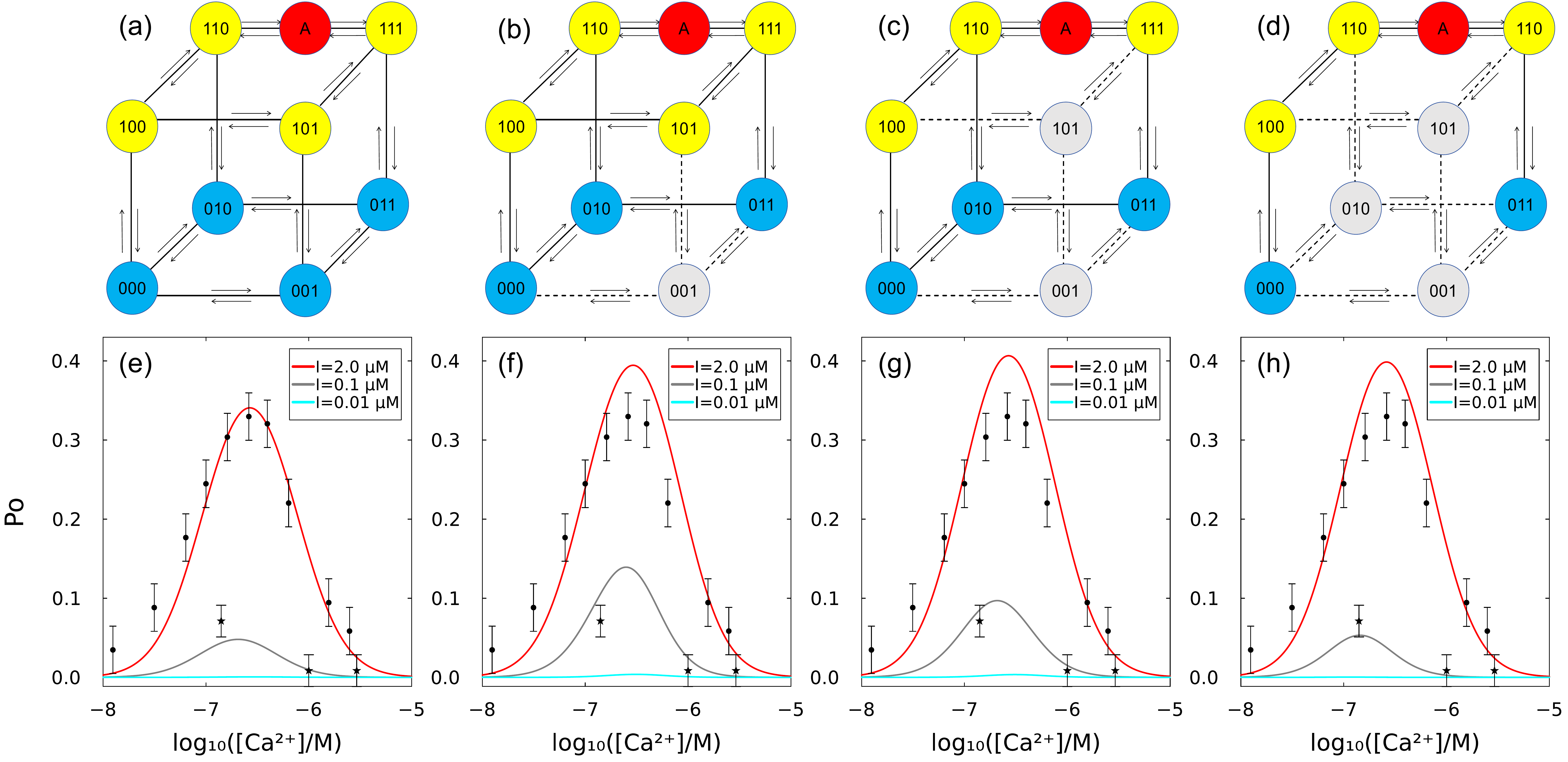}
\caption{Comparison of the nine-state model and three simplified models obtained
by sequentially removing the 001, 101, and 010 states (gray indicates removed
states). The lower panel shows the simulated open probability ($P_{\rm O}$) for
each model alongside experimental data from single-channel patch-clamp
recordings of IP$_3$R on the native nuclear membrane. Circles represent data for
[IP$_3$] = 0.2 $\mu$M, while stars indicate data for [IP$_3$] = 0.01
$\mu$M~\cite{mak_spontaneous_2003,tu_functional_2003}.  }
\label{fig: role}
\end{figure*}

The results, presented in Fig.~\ref{fig: role}, demonstrate that all models
retain a bell-shaped $P_{\rm O}$ curve, indicating that the fundamental
dependence of IP\textsubscript{3}R activity on $[\text{Ca}^{2+}]$ remains intact
despite the removal of specific states. This observation is consistent with the
findings of Shuai et al.~\cite{shuai_investigation_2009}, who reported that the
001 and 101 states could be omitted without significantly altering the channel's
functional behavior. Notably, the full nine-state model provides predictions
that closely match experimental
data~\cite{mak_spontaneous_2003,tu_functional_2003}, underscoring its ability to
accurately capture the Ca\textsuperscript{2+}-dependent regulation of
IP\textsubscript{3}Rs.

The overall bell-shaped dependence of $P_{\rm O}$ on $[\text{Ca}^{2+}]$ remains
largely unchanged across different models. At IP$_3$ concertraiton $[\text{IP}_3]$ = 2.0~$\mu$M, the
nine-state model predicts a peak $P_{\rm O}$ of 0.35 at $[\text{Ca}^{2+}] =
10^{-6.6}~$M. Minor variations are observed in peak $P_{\rm O}$ when modifying
the model, with the removal of the 001 state slightly increasing the maximum
$P_{\rm O}$ to 0.4. Additional adjustments involving the 101 and 010 states
result in negligible further changes, suggesting that the fundamental channel
behavior remains intact, with only minor contributions from these states at high
IP\textsubscript{3} concentrations.

At lower IP\textsubscript{3} concentrations ($[\text{IP}_3]$ = 0.1~$\mu$M), the effect of the 001 state is slightly more noticeable, with the full nine-state model predicting a peak $P_{\rm O}$ of 0.05. The absence of the 001 state shifts the peak of the $P_{\rm O}$ curve toward a lower $[\text{Ca}^{2+}]$ concentration, indicating a mild influence on calcium sensitivity. However, the removal of the 101 and 010 states introduces little additional change, reaffirming that the overall regulatory dynamics are preserved across models.

At very low IP\textsubscript{3} concentrations ($[\text{IP}_3]$ = 0.01~$\mu$M), all models predict an opening probability near zero, indicating that such low IP\textsubscript{3} levels are insufficient to activate the channel. This result aligns with experimental observations \cite{alzayady_defining_2016}, further confirming the model’s validity. Under these minimal IP\textsubscript{3} conditions, differences between models become negligible, underscoring the dominant role of IP\textsubscript{3} binding in regulating channel activity.

\subsection{Calcium oscillation}

\begin{figure*}[htbp]
\includegraphics[bb=0 0 960 400, clip, scale=0.555]{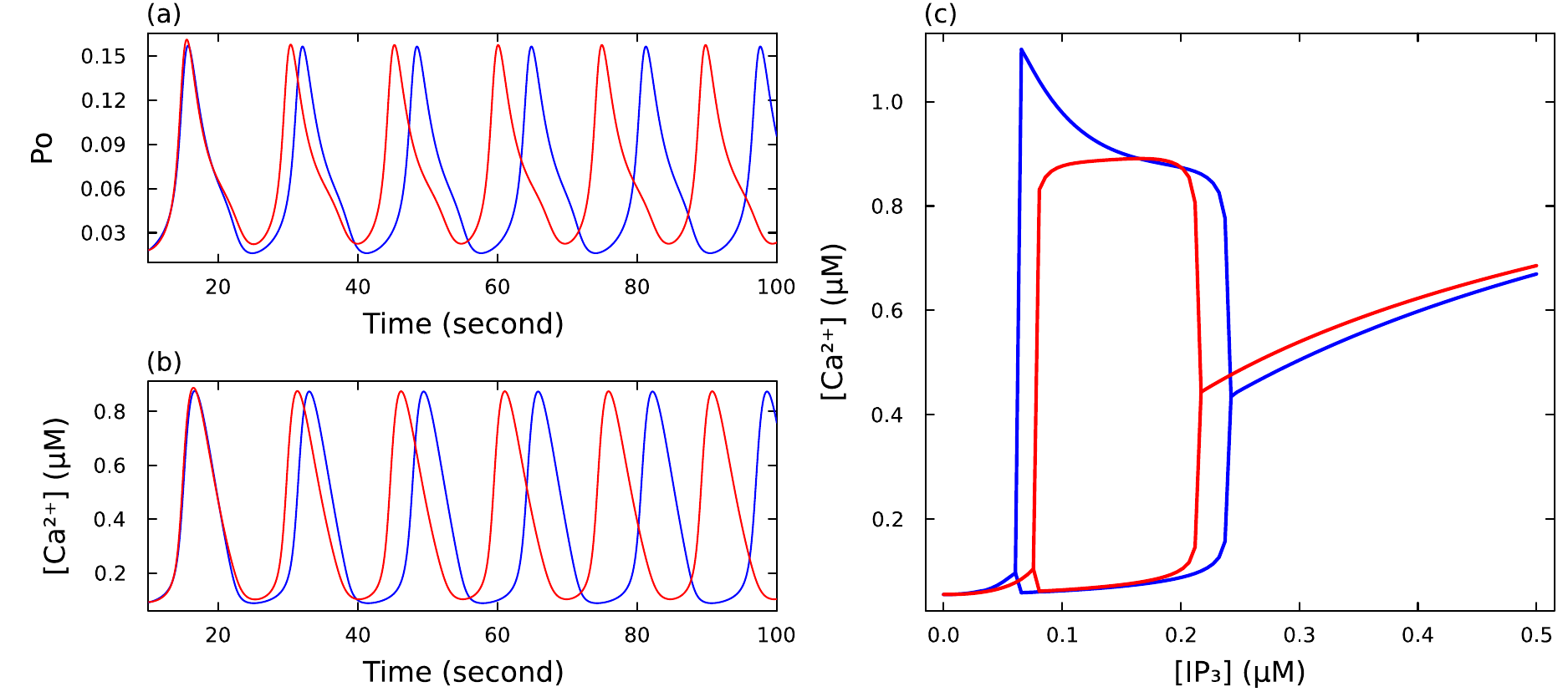}
\caption{Calcium oscillations. (a) Variation in open probability over time. (b) Variation in calcium concentration $[\text{Ca}^{2+}]$ over time. (c) Bifurcation diagram of calcium concentration as a function of [IP$_3$] levels. In all panels, the red line represents the nine-state model and the blue line represents the six-state model.}
\label{fig:Calcium oscillation}
\end{figure*}

Using both the nine-state and simplified six-state models, we simulated calcium
oscillations by incorporating the differential equations governing calcium
concentration dynamics in the cytoplasm and ER, as described in Eq.~(\ref{de}).
Under an IP\textsubscript{3} concentration of 0.2~$\mu$M, the models
successfully reproduced calcium oscillations. The results in
Fig.~\ref{fig:Calcium oscillation} (a) and (b) show that the oscillation period
in the cytoplasm obtained with the nine-state model is approximately 14.8
seconds, matching experimentally observed oscillation periods in non-excitable
cells \cite{parekh2011decoding,Berridge_1989}. The maximum open probability,
$P_{\rm O}$, reaches approximately 0.16. When using the six-state model, the
oscillation period increases slightly, while both the oscillation amplitude and
$P_{\rm O}$ remain nearly unchanged. Further testing, by turning off additional
states, confirmed that these states are essential for reproducing the oscillatory
behavior.

To provide a more comprehensive picture of calcium oscillations in our model, we
present bifurcation diagrams for [IP\textsubscript{3}] and
[Ca\textsuperscript{2+}] derived from both the nine-state and six-state models
in Fig.\ref{fig:Calcium oscillation} (c). Unlike the DYK model, our model
demonstrates increased sensitivity to [IP\textsubscript{3}], enabling
oscillations to occur at lower IP\textsubscript{3} concentrations. In the
nine-state model, calcium oscillations first appear when the
[IP\textsubscript{3}] concentration exceeds 0.08$\mu$M. These oscillations
remain stable within the range of 0.1-0.2~$\mu$M, maintaining a relatively
constant amplitude. However, as [IP\textsubscript{3}] increases further, the
oscillation amplitude abruptly vanishes when [IP\textsubscript{3}] approaches
0.22~$\mu$M. Although the six-state model begins to exhibit oscillations at a slightly lower [IP\textsubscript{3}] threshold of 0.065~$\mu$M—where [Ca\textsuperscript{2+}] peaks at 1.10~$\mu$M—its overall dynamic behavior remains highly consistent with that of the full nine-state model.

While both the original DYK model with eight states and our model display an
[IP\textsubscript{3}] window for calcium oscillations, a key distinction lies in
our model, which incorporates pre-activated states. In our
model, the oscillation amplitude shows a sharp, abrupt increase upon surpassing
the [IP\textsubscript{3}] threshold. This increase occurs immediately, reaching
its maximum value and remaining nearly constant for a brief period before
diminishing. In contrast, the DYK model exhibits a more gradual increase in
oscillation amplitude, with the peak occurring only after an
[IP\textsubscript{3}] concentration increase of approximately
0.2~$\mu$M~\cite{de_young_single-pool_1992}. These results suggest that a
critical [IP\textsubscript{3}] threshold exists, beyond which calcium binding is
significantly accelerated. This rapid binding leads to the swift opening of
multiple IP\textsubscript{3}R channels and the initiation of calcium
oscillations. This mechanism emphasizes IP\textsubscript{3}'s ability to
regulate calcium oscillations with high precision, ensuring that the
oscillations can be finely controlled and stabilized within the system.

\section{Summary and discussion}\label{sec4}

In this work, we developed a novel nine-state IP\textsubscript{3}R channel model
incorporating pre-activated states, aligning it with recent electron microscopy
findings. We calculated the distribution of IP\textsubscript{3}R states under
saturating IP\textsubscript{3} conditions and found the results comparable to
the electron microscopy observations. Incorporating calcium exchange
between the cytosol and the ER enables the reproduction of calcium oscillations, and is well
consistent with the experimental result.

Our model replicates key features of Ca\textsuperscript{2+}-dependent
IP\textsubscript{3}R activation and inhibition observed in previous biochemical
and electrophysiological experiments~\cite{foskett_inositol_2007}. It reproduces
similar trends in free Ca\textsuperscript{2+} levels across four selected states
under saturating IP\textsubscript{3} conditions, notably capturing the
bell-shaped dependence for both the preactivated and activated states. Although
our results are qualitatively consistent with electron microscopy
data~\cite{paknejad_structural_2023}, quantitative discrepancies persist.
Structural studies suggest that some receptor subunits may remain unbound due to
incomplete ligand occupancy or conformational heterogeneity~\cite{Fan_2022}.
Future refinements, such as incorporating partial IP\textsubscript{3} occupancy
or cooperative interactions between subunits, could help resolve these
differences.

We also disuss the role of the states in the model. By removing the 001, 101, and 010 states, we found
that the bell-shaped relationship between open probability (P\textsubscript{O})
and calcium concentration remained consistent, with the nine-state model closely
matching experimental data. Although the six-state reduced model requires slightly lower IP\textsubscript{3}
concentrations to trigger calcium oscillations, it maintains similar periods and
amplitudes. Such results suggest that the 001, 101, and 010 states may play minor roles 
in calcium signaling, at least in the case considered in the current work.

In our model, a key characteristic is that the calcium oscillation sharply initiates as the IP$_3$ concentration rises, remaining at a constant amplitude during its course. Once it reaches a certain level, the oscillation abruptly ceases. This indicates that IP$_3$ exerts a switch-like regulatory mechanism over the calcium oscillation, where the system displays an all-or-none response: no oscillation below a threshold, sustained oscillation with fixed amplitude above it, and termination beyond a higher threshold. This mechanism is crucial for cells that rely on binary signaling to regulate downstream processes.

A key contributor to the switch-like behavior observed in our simulations is the presence of a pre-activated intermediate state in the IP$_3$R gating scheme. Cryo-EM structures of IP$_3$R1 have revealed partially shifted conformations—such as contraction of the ARM2 domain—that are consistent with a primed, non-conductive state preceding full channel opening~\cite{paknejad_structural_2023}. In our model, elevated IP$_3$ stabilizes this intermediate conformation, resulting in a substantial accumulation of subunits in this state prior to Ca$^{2+}$-dependent activation.
 
Because this intermediate lies energetically closer to the open state than the fully closed conformation, it reduces the activation barrier and shortens the latency between Ca$^{2+}$ binding and channel opening. As a result, a modest increase in cytosolic Ca$^{2+}$ can rapidly trigger the transition of many receptors from the intermediate to the open state, producing the steep, threshold-like Ca$^{2+}$ release characteristic of Fig.~\ref{fig:Calcium oscillation}. 

In physiological contexts requiring precise control over calcium dynamics—such as neuronal signaling, immune cell activation, and hormone secretion—this mechanism ensures a rapid and coordinated cellular response. The constant oscillation amplitude observed after crossing the threshold can be interpreted as a frequency-encoded signaling mechanism, where the information is transmitted via oscillation frequency rather than amplitude variation. This mode of encoding is well-established in various calcium-dependent pathways, including transcriptional regulation and enzyme activation~\cite{Simms_2014,berridge2000calcium}. The presence of a pre-activated intermediate state thus provides a structural and dynamic basis for reliable on–off switching, stable amplitude generation, and frequency modulation, offering both precision and flexibility in cellular calcium signaling.

In conclusion, our
findings highlight the importance of pre-activated state in shaping
threshold-dependent behavior, suggesting that the state are a fundamental
feature of IP\textsubscript{3}R dynamics, enabling precise control over
calcium-mediated cellular functions. It is vital for calcium signaling, allowing cells to rapidly switch
between non-oscillatory and oscillatory states. These dynamics are crucial for
various calcium-dependent processes, such as transcription regulation and enzyme
activation, highlighting the model's potential in understanding
frequency-encoded signaling in cellular functions.

\begin{acknowledgments}
  This project is supported by the National Natural Science
  Foundation of China (Grant No. 12475080).
  \end{acknowledgments}

\bibliographystyle{apsrev4-2}
\bibliography{Ca} 
\end{document}